\documentclass[aps,prb,twocolumn,groupaddress,showpacs,floatfix]{revtex4}
\usepackage{graphicx}
\usepackage{dcolumn}
\usepackage{bm}
\usepackage{stmaryrd}
\usepackage{latexsym}
\usepackage{amssymb}
\usepackage{amsfonts}
\usepackage{amsmath}

\begin{document}
\renewcommand{\thefigure}{\arabic{figure}}

\title{Density-functional theory of inhomogeneous electron systems in thin quantum wires}
\author{Saeed H. Abedinpour}
\author{Marco Polini}
\email{m.polini@sns.it}
\author{Gao Xianlong}
\author{M.P. Tosi}
\affiliation{NEST-CNR-INFM and Scuola Normale Superiore, I-56126 Pisa, Italy}
\date{\today}

\begin{abstract}
Motivated by current interest in strongly correlated quasi-one-dimensional ($1D$) Luttinger liquids subject to axial confinement, we present a novel density-functional study of few-electron systems confined by power-low external potentials inside a short portion of a thin quantum wire. The theory employs the 
$1D$ homogeneous Coulomb liquid as the reference system for a Kohn-Sham treatment and transfers the Luttinger ground-state correlations to the inhomogeneous electron system by means of a suitable local-density approximation (LDA) to the exchange-correlation energy functional. We show that such $1D$-adapted LDA is appropriate for fluid-like states at weak coupling, but fails to account for the transition to a ``Wigner molecules'' regime of electron localization as observed in thin quantum wires at very strong coupling. A detailed analyzes is given for the two-electron problem under axial harmonic confinement.
\end{abstract}
\maketitle

\section{Introduction}
\label{sect:intro}

One-dimensional ($1D$) quantum many-body systems of interacting fermions have attracted theoretical and experimental interest for more than fifty years~\cite{giamarchi_book}. Contrary to what happens in higher dimensionality, these systems cannot be described by the conventional Landau theory of normal Fermi liquids~\cite{Giuliani_and_Vignale} due to a subtle interplay between topology and interactions. The appropriate paradigm for $1D$ interacting fermions is instead provided by the Luttinger-liquid concept introduced by Haldane in the early eighties~\cite{haldane}.

Strongly correlated $1D$ systems that are nowadays available for experiment range from ultra-cold atomic gases~\cite{reviews} to electrons in single-wall carbon nanotubes~\cite{Saito_book} and in semiconductor quantum wires~\cite{auslaender_2002,auslaender_2005}. Chiral Luttinger liquids at fractional quantum-Hall edges~\cite{xLL} also provide an example of $1D$ electronic conductors and have been the subject of intense experimental and theoretical studies~\cite{xLL_experiments, giovanni_allan}. In many experimental situations the translational invariance of the fluid is broken by the presence of inhomogeneous external fields. Examples are the confining potential provided by magnetic and optical traps for ultra-cold gases~\cite{reviews} and the barriers at the end of a quantum wire segment in cleaved edge overgrowth samples~\cite{auslaender_2005}. These strong perturbations induce the appearance of a new length scale and can cause novel physical behaviors relative to the corresponding unperturbed, Galileian-invariant model system.

Of special relevance to the present work are the studies carried out in Ref.~\onlinecite{auslaender_2005}, where momentum-resolved tunneling experiments between two closely situated parallel quantum wires have been carried out to probe the phenomenon of spin-charge separation in a Luttinger liquid~\cite{michele_2002}. In these experiments a top gate is used to deplete the central portion of one of the two wires, thus locally decreasing the electron density, and a dramatic transition is observed when the electron density is reduced below a critical value. There is strong evidence that in this regime the electrons in the depleted wire segment are separated by barriers from the rest of the wire~\cite{fiete_2005}, and it is suggested that the electrons in the segment are localized by the combined effect of the barriers and of the electron-electron interactions. The magnetic-field dependence of the tunneling conductance for a field perpendicular to the plane of the wires provides a  direct probe of the many-body wavefunction of the localized electrons~\cite{fiete_2005,mueller_2005}, offering the possibility to investigate systematically the role of interactions in creating exotic phases of matter in reduced dimensionality. In fact, the experimental parameters in Ref.~\onlinecite{auslaender_2005} are such that the electrons in the wire segment are in the strong-coupling regime. An exact diagonalization study was carried out in Ref.~\onlinecite{fiete_2005} for a number $N$ of electrons up to $4$.

A powerful theoretical tool to study the interplay between interactions and inhomogeneity from external fields of arbitrary shape is density-functional theory (DFT), based on the Hohenberg-Kohn theorem and the Kohn-Sham mapping~\cite{dft}. Many-body effects enter DFT {\it via} the exchange-correlation (xc) functional, which is often treated by a local-density approximation (LDA) requiring as input the xc energy of a homogeneous  reference fluid. For $2D$ and $3D$ electronic systems the underlying reference fluid usually is the homogeneous electron liquid (EL), whose xc energy is known to a high degree of numerical precision from quantum Monte Carlo (QMC) studies~\cite{qmc_3D&2D}. Several density-functional schemes have also been proposed for strongly correlated $1D$ systems~\cite{soft,lima_prl_2003,burke_2004,kim_2004,gao_prb_2006} and in the case of the $1D$ Luttinger liquid with repulsive contact interactions, where the xc energy of the homogeneous fluid is exactly known from Bethe-{\it Ansatz} solutions, tests of the LDA have been carried out against QMC data~\cite{gao_prb_2006}.

In the present work we test a novel LDA xc functional to treat few-electron systems confined by power-law potentials inside a segment of a quantum wire. The homogeneous reference system that we adopt is the $1D$ EL with Coulomb interactions, previously studied by a number of authors~\cite{1deg,hausler,fogler} and very recently evaluated by a novel lattice-regularized Diffusion Monte Carlo method~\cite{casula_2006}. 
The correlation energy determined in this latter study is used in our LDA calculations, whereas earlier DFT-based studies of $1D$ ``quantum dots''~\cite{reiman_1998} have used the correlation energy of a $2D$ EL. While such choice can be justified for thick wires, a $1D$ reference fluid is more appropriate to treat inhomogeneous electron systems in ultrathin wires of our present interest (see also the discussion given in Ref.~\onlinecite{fogler}). 
We nevertheless find that the $1D$-adopted LDA is unable to describe the transition of the confined electrons from a fluid-like state to the localized ``Wigner-like'' state that is observed to occur as the coupling strength is increased. In essence, the confining potential pins the phase of density oscillations in much the same way as an impurity inserted into the infinitely extended $1D$ fluid does in producing Friedel oscillations in the surrounding electron density. However, a cross-over from a $2k_{\rm F}$ to a $4k_{\rm F}$ periodicity occurs in these oscillations with increasing coupling in the Luttinger liquid. 
We proceed in the later part of the paper to give a detailed analysis of this transition in the case of two electrons subject to axial harmonic confinement in a wire segment. From previous work on the two-particle problem with contact repulsive interactions~\cite{saeed_pra_2006} we presume that a local spin-density approximation could help in transcending the limitations of the LDA. 

The outline of the paper is briefly as follows. In Sect.~\ref{sect:theory} we introduce the Hamiltonian that we use for the system of present interest, and in Sect.~\ref{sect:dft} we describe our self-consistent DFT approach and the LDA that we employ for the xc potential. In Sect.~\ref{sect:numerical_results} we report and discuss our main results for the fluid state at weak coupling, while in Sect.~\ref{sect:strong_coupling} we focus on the two-electron problem. Finally, Sect.~\ref{sect:discussion_conclusions} summarizes our main conclusions.

\section{The model}
\label{sect:theory}

We consider $N$ electrons of band mass $m$ confined inside an axially symmetric quantum wire. 
The transverse confinement is provided by a tight harmonic potential with angular frequency $\omega_\perp$, 
\begin{equation}\label{eq:transverse_oscillator}
V_\perp(x,y)=\frac{1}{2} m \omega^2_\perp (x^2+y^2)\,.
\end{equation}
The electrons are also subject to a longitudinal potential $V_{\rm ext}(z)$ along the wire axis. 
In the $1D$ limit (see below) the transverse motion can be taken as frozen into the ground state of the $2D$ oscillator, $\varphi({\bf r}_\perp)=(2 \pi b^2)^{-1/2}\exp{[-{\bf r}^2_\perp/(4 b^2)]}$ with $b^2=\hbar/(2 m\omega_\perp)$. 
The parameter $b$ thus measures the transverse wire radius. On integrating out the transverse degrees of freedom one ends up with the effective $1D$ Hamiltonian
\begin{equation}\label{eq:1deg}
{\cal H}=-\frac{\hbar^2}{2m}\sum_{i}\frac{\partial^2}{\partial z^2_i}+\frac{1}{2}\sum_{i \neq j}v_b(|z_i-z_j|)+\sum_i V_{\rm ext}(z_i)\,,
\end{equation}
where
\begin{equation}\label{eq:vb-1d}
v_b(z)=\frac{\sqrt{\pi}}{2}\frac{e^2}{\kappa b}\exp{[z^2/(4b^2)]}{\rm erfc}[z/(2b)]
\end{equation}
is the renormalized interelectron potential~\cite{hausler}. Here $\kappa$ is a background dielectric constant and ${\rm erfc}(x)$ is the complementary error function~\cite{abram}. It is easy to check that the potential in Eq.~(\ref{eq:vb-1d}) becomes purely Coulombic at large distance~\cite{footnote}, $v_b(z)\rightarrow e^2/(\kappa |z|)$ for $|z|\rightarrow \infty$. At zero interelectron separation the electron-electron potential goes to a positive constant. Equation (\ref{eq:vb-1d}) yields a linear approach to a constant, the cusp being an artifact of wavefunction factorization~\cite{hausler}.

The last term in Eq.~(\ref{eq:1deg}) gives the coupling of the electrons to the axial external potential and, following Tserkovnyak {\it et al.}\cite{auslaender_2002}, we consider power-law potentials of the type
\begin{equation}\label{eq:external_potential}
V_{\rm ext}(z)=V_{\beta}|z|^\beta
\end{equation}
with $\beta \geq 2$ and $V_\beta=2^{\beta+1}\hbar^2/(m L^{2+\beta})$. For $\beta=2$ the confinement is harmonic, 
$V_{\rm ext}(z)=m \omega^2_{\|} z^2/2$ with angular frequency $\omega_{\|}=4\hbar/(m L^2)$, while $V_{\rm ext}(z)$ becomes a square well of size $L$ in the limit $\beta\rightarrow +\infty$.

Choosing $L/2$ as the unit of length and $2\hbar^2/(m L^2)$ as the unit of energy, the Hamiltonian becomes
\begin{equation}\label{eq:1deg_dimensionless}
{\cal H}=-\sum_{i}\frac{\partial^2}{\partial x^2_i}+\frac{\lambda}{2}\sum_{i \neq j}{\cal F}(|x_i-x_j|)+\sum_i\,|x_i|^\beta
\end{equation}
with $x=2z/L$, $\lambda=\sqrt{\pi} {\bar L}^2/(4 {\bar b})$ and ${\cal F}(x)=\exp{[{\bar L}^2 x^2/(16 {\bar b}^2)]}{\rm erfc}[{\bar L} x/(4{\bar b})]$. Here 
${\bar L}=L/a_{\rm B}$ and ${\bar b}=b/a_{\rm B}$, $a_{\rm B}=\hbar^2\kappa/(m e^2)$ being the effective Bohr radius. We see from 
Eq.~(\ref{eq:1deg_dimensionless}) that the physical properties of the system are determined by the four dimensionless parameters $N, {\bar b}, \beta$, 
and ${\bar L}$. Note that while ${\cal F}(x)$ is controlled only by the ratio ${\bar L}/{\bar b}$, the parameter $\lambda$ contains two powers of ${\bar L}$ and one power of ${\bar b}$. Electron-electron interactions are expected to become dominant in ultrathin wires with ${\bar b} \lesssim 1$ 
and for weak confinements (${\bar L}\gg 1$). In the experiments of Ref.~\onlinecite{auslaender_2005} (with $a_B \simeq 9.8~{\rm nm}$ for GaAs) ${\bar b}\approx 1$ and ${\bar L} \approx 100$, so that the electrons are in a strong-coupling regime ($\lambda\approx 4\times 10^{3}$). In fact, the electron-electron coupling is also influenced by the exponent $\beta$, which determines the spill-out of the electron density and hence the system diluteness. For given ${\bar b}$ and ${\bar L}$, harder boundaries (larger $\beta$) imply a more efficient confinement ({\it i.e.} higher average density) and thus reduce the role of the many-body interactions.

\section{Density-functional approach}
\label{sect:dft}

Within the Kohn-Sham version of 
DFT the ground-state density $n_{\rm GS}(z)$ is calculated by self-consistently solving the Kohn-Sham equations for single-particle orbitals $\varphi_\alpha(z)$,
\begin{equation}\label{eq:kss}
\left[-\frac{\hbar^2}{2m}\frac{d^2}{d z^2}+V_{\rm KS}[n_{\rm GS}](z)\right]\varphi_\alpha(z)=\varepsilon_\alpha\varphi_\alpha(z)
\end{equation}
with $V_{\rm KS}(z)=v_{\rm H}(z)+v_{\rm xc}(z)+V_{\rm ext}(z)$, together with the closure
\begin{equation}\label{eq:closure}
n_{\rm GS}(z)=\sum_{\alpha}\Gamma_\alpha\left|\varphi_\alpha(z)\right|^2\,.
\end{equation}
Here the sum runs over the occupied orbitals and the degeneracy factors $\Gamma_\alpha$ satisfy the sum rule 
$\sum_\alpha \Gamma_\alpha=N$. The first term in the effective Kohn-Sham potential is the Hartree term
\begin{equation}\label{eq:hartree_potential}
v_{\rm H}[n_{\rm GS}](z)=\int_{-\infty}^{+\infty}dz' v_b(|z-z'|)n_{\rm GS}(z')\,,
\end{equation} 
while the second term is the xc potential, defined as the functional derivative of the xc energy $E_{\rm xc}[n]$ evaluated at the ground-state density profile, $v_{\rm xc}=\delta E_{\rm xc}[n]/\delta n(z)|_{\rm GS}$. The total ground-state energy of the system is given by 
\begin{eqnarray}
E_{\rm GS}&=&\sum_\alpha\Gamma_\alpha\varepsilon_\alpha-\int_{-\infty}^{+\infty}dz\,
v_{\rm xc}[n_{\rm GS}](z)n_{\rm GS}(z)\nonumber\\
&-&\frac{1}{2}\int_{-\infty}^{+\infty}dz\int_{-\infty}^{+\infty}dz' v_b(|z-z'|)n_{\rm GS}(z)n_{\rm GS}(z')\nonumber\\
&+&E_{\rm xc}[n_{\rm GS}]\,.
\end{eqnarray}
Equations~(\ref{eq:kss}) and~(\ref{eq:closure}) provide a formally exact scheme to calculate $n_{\rm GS}(z)$ and $E_{\rm GS}$, but $E_{\rm xc}$ and $v_{\rm xc}$ need to be approximated.

As mentioned above in Sect.~\ref{sect:intro}, in this work we have chosen the $1D$ EL, described by the Hamiltonian (\ref{eq:1deg}) with $V_{\rm ext}(z)=0$, as the homogeneous reference fluid. In the thermodynamic limit and in the absence of spin polarization this model is described by two dimensionless parameters only, $r_s$ and ${\bar b}$. Here $r_s=(2 n a_{\rm B})^{-1}$ is the usual Wigner-Seitz dimensionless parameter, defined in terms of the average $1D$ density $n$. We adopt the LDA functional
\begin{eqnarray}\label{eq:lda_energy}
E_{\rm xc}[n] \rightarrow  E^{\rm LDA}_{\rm xc}[n] = \int_{-\infty}^{+\infty}dz n(z) \varepsilon^{\rm hom}_{\rm xc}(r_s(z))
\end{eqnarray}
with $r_s(z)=[2 n_{\rm GS}(z) a_{\rm B}]^{-1}$ and $\varepsilon^{\rm hom}_{\rm xc}(r_s)=\varepsilon^{\rm hom}_{\rm x}(r_s)+\varepsilon^{\rm hom}_{\rm c}(r_s)$. The exchange energy $\varepsilon^{\rm hom}_{\rm x}$ of the $1D$ EL (per particle)  is calculated from
\begin{equation}\label{eq:exchange_energy}
\varepsilon^{\rm hom}_{\rm x}(r_s)=\frac{1}{2}\int_{-\infty}^{+\infty} \frac{dq}{2\pi} v_b(q)[S_0(q)-1]\,,
\end{equation}
where $v_b(q)=(e^2/\kappa)\exp{(q^2b^2)}{\rm E}_1(q^2b^2)$ is the Fourier transform of the interaction potential, with ${\rm E}_1(x)$ being the exponential integral~\cite{abram}, and $S_0(q)$ is the noninteracting-gas structure factor ($S_0(q)=q/(2k_{\rm F})$ for $q\leq 2k_{\rm F}$ and $1$ elsewhere). The correlation energy $\varepsilon^{\rm hom}_{\rm c}$ determined by Casula {\it et al.}~\cite{casula_2006} is given by the parametrization formula
\begin{equation}\label{eq:corr_energy}
\varepsilon^{\rm hom}_{\rm c}(r_s)=-\frac{r_s}{A+Br^{\gamma}_s+Cr^2_s}\ln{(1+ D r_s+ E r^{\gamma '}_s)}\,,
\end{equation}
in units of the effective Rydberg $e^2/(2\kappa a_{\rm B})$. The values of the seven parameters  in this expression are reported in Table IV of Ref.~\onlinecite{casula_2006} for several values of ${\bar b}$ in the range $0.1\leq {\bar b} \leq 4$. As discussed in Ref.~\onlinecite{casula_2006}, Eq.~(\ref{eq:corr_energy}) incorporates the exactly-known weak-coupling limit ($r_s\rightarrow 0$) and fits very well their numerical data in the range $0.05 \leq  r_s \leq 50$. Finally, the LDA xc potential is calculated from Eq.~(\ref{eq:lda_energy}) as
\begin{widetext}
\begin{equation}\label{eq:xc_potential}
v^{\rm LDA}_{\rm xc}[n_{\rm GS}](z)=\left.\frac{\delta E^{\rm LDA}_{\rm xc}[n]}{\delta n}\right|_{\rm GS}=
\left. \left(1-r_s\frac{\partial}{\partial r_s}\right)\varepsilon^{\rm hom}_{\rm xc}(r_s)\right|_{r_s\rightarrow [2n_{\rm GS}(z)a_{\rm B}]^{-1}}\,.
\end{equation}
\end{widetext}
We have calculated numerically the derivative of the exchange energy as
\begin{eqnarray}\label{eq:derivative_exchange}
\frac{\partial \varepsilon^{\rm hom}_{\rm x}(r_s)}{\partial r_s}
&=&-\frac{1}{2 r^2_s a_{\rm B}}\int_{0}^{1} d{\bar q}~({\bar q}-1)v_b({\bar q}) \nonumber \\
&+&\frac{1}{2r_s a_{\rm B}}\int_{0}^{1} d{\bar q}~({\bar q}-1)\frac{\partial v_b({\bar q})}{\partial r_s}\,,
\end{eqnarray}
where ${\bar q}=q/(2k_{\rm F})$. Notice that $v_b({\bar q})$ is $r_s$-dependent.

\section{Numerical results for the fluid state}
\label{sect:numerical_results}

\begin{figure}
\begin{center}
\includegraphics[width=0.80\linewidth]{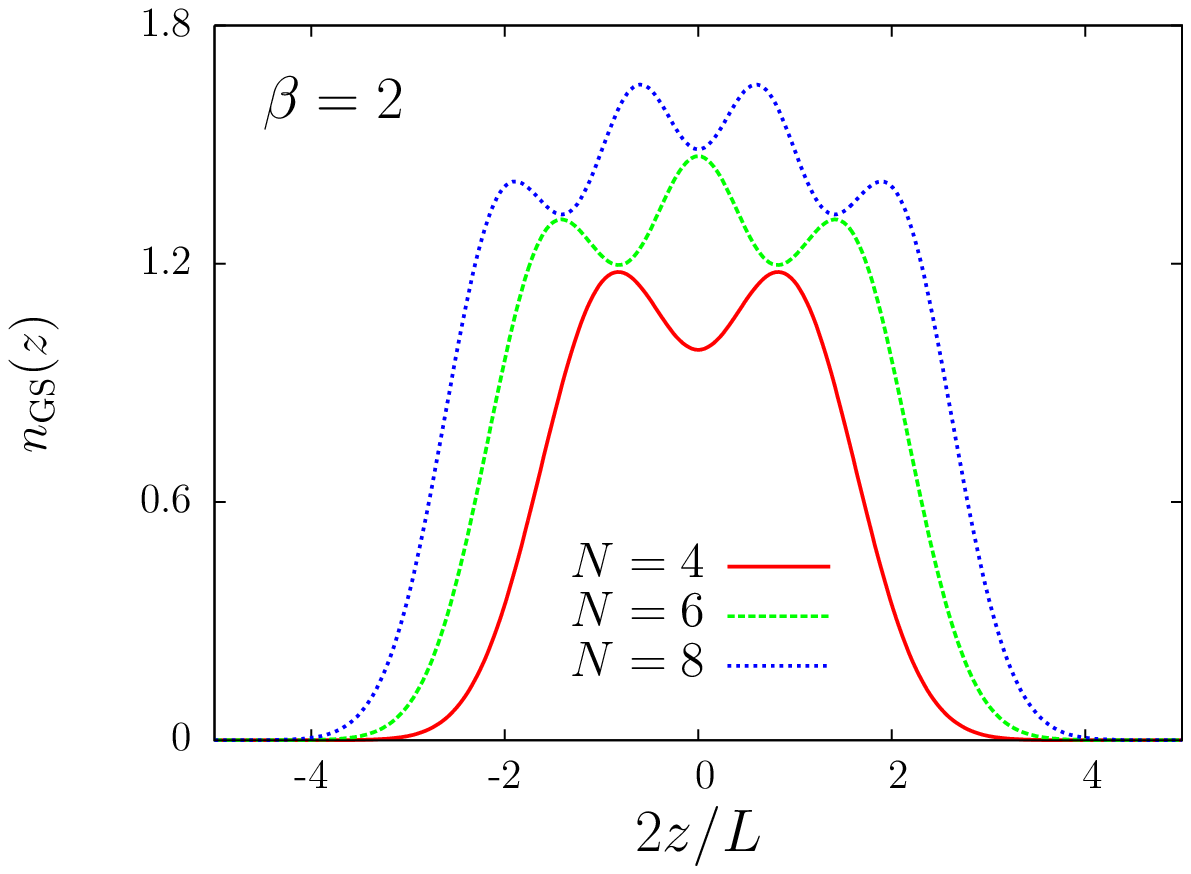}\\
\includegraphics[width=0.80\linewidth]{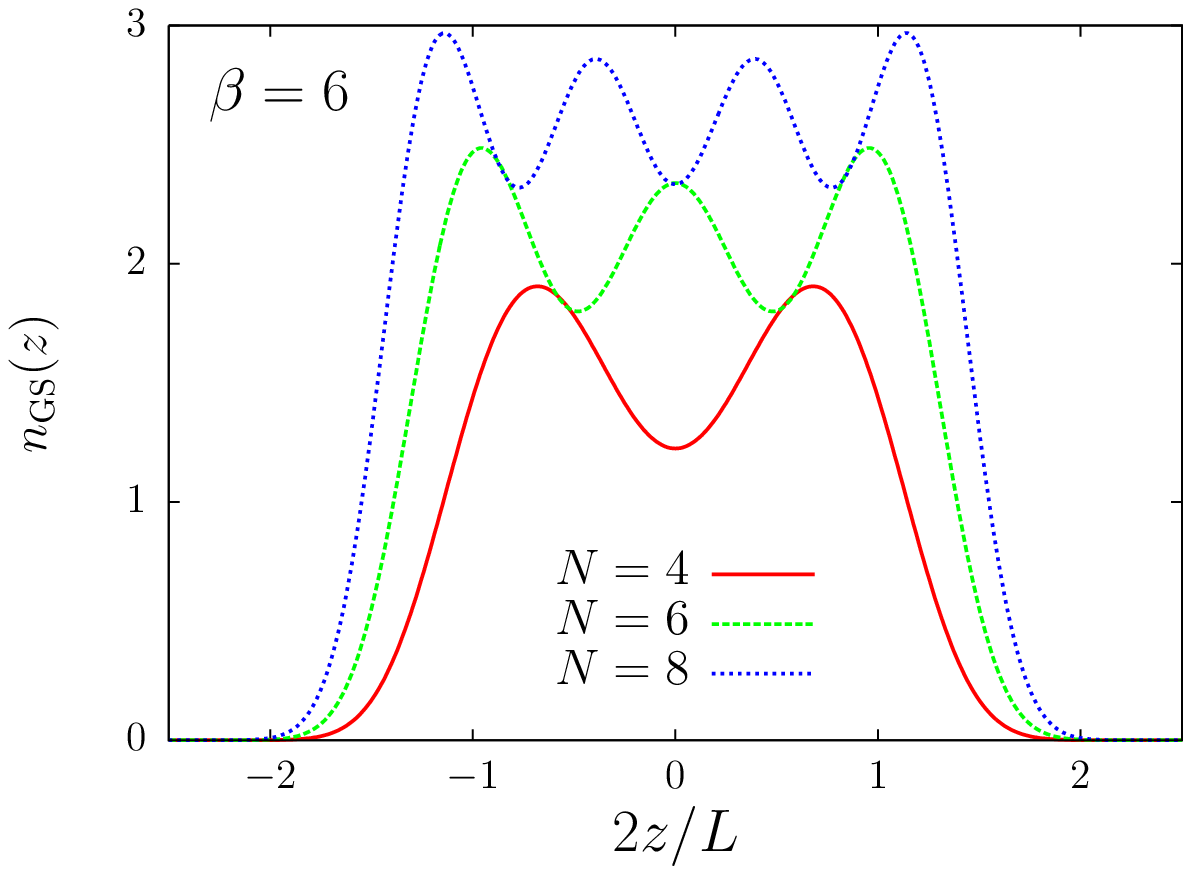}
\caption{(Color online) Top panel: Density profile $n_{\rm GS}(z)$ (in units of $2/L$) as a function of $2z/L$ 
for $N=4, 6$, and $8$ electrons confined by an external potential with $\beta=2$ and $L=a_{\rm B}$ in a thin wire of radius $b=0.1 a_{\rm B}$. 
Bottom panel: Same as in the top panel but for $\beta=6$.\label{fig:one}}
\end{center}
\end{figure}

\begin{figure}
\begin{center}
\includegraphics[width=0.80\linewidth]{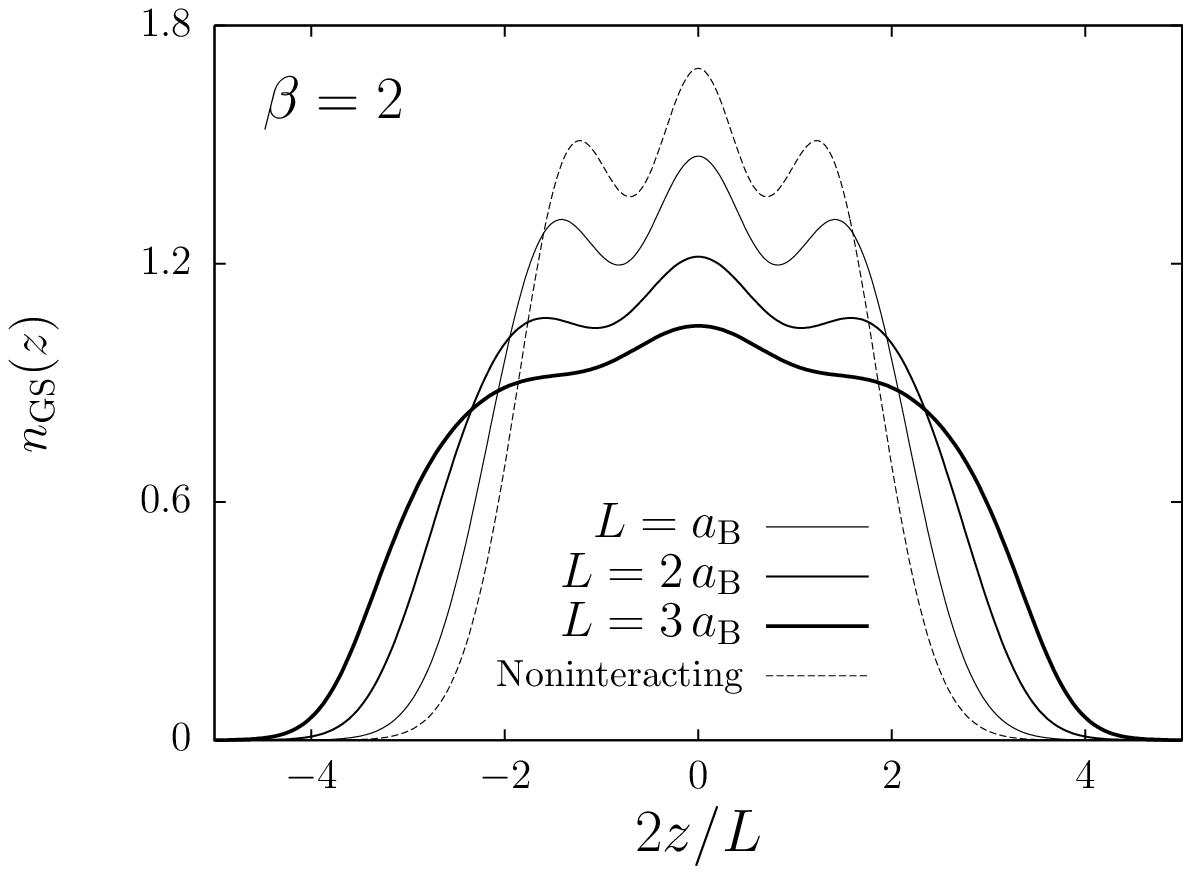}\\
\includegraphics[width=0.80\linewidth]{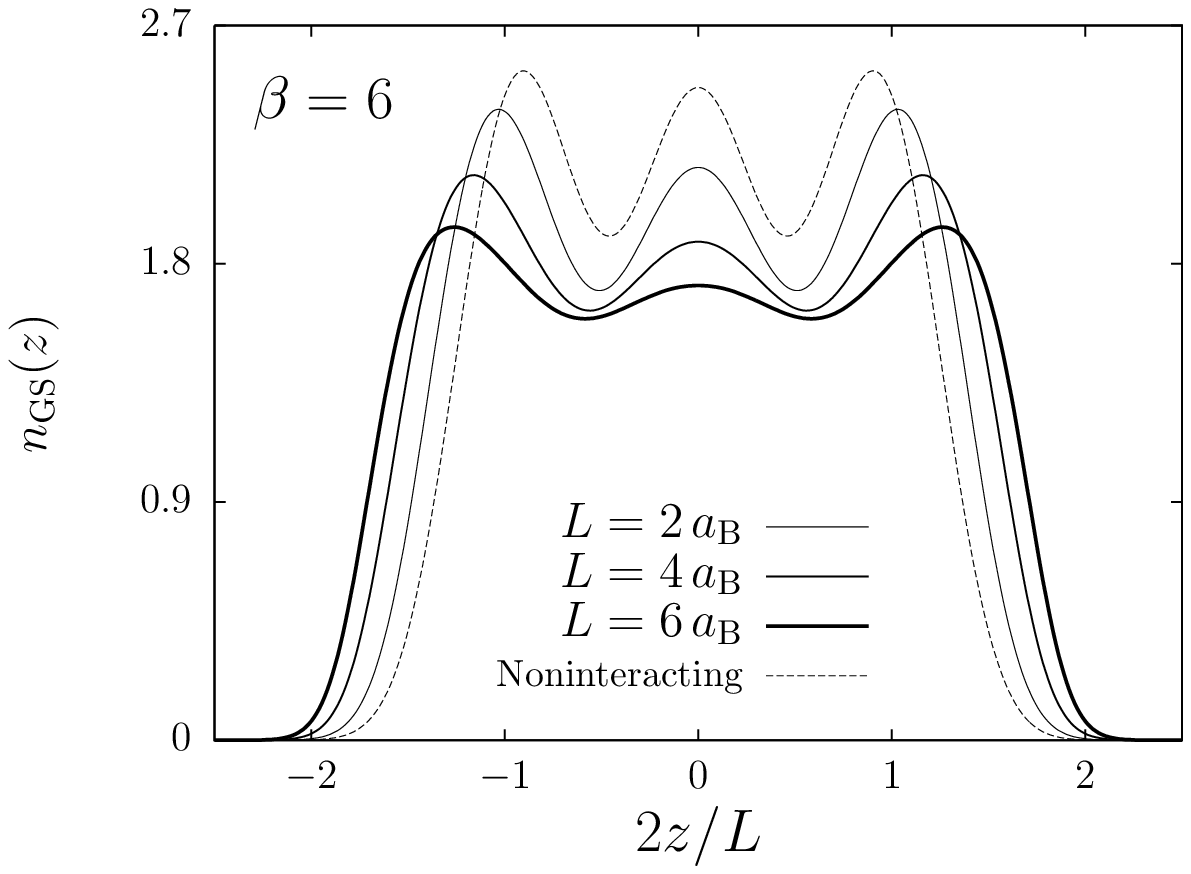}
\caption{Top panel: Density profile $n_{\rm GS}(z)$ (in units of $2/L$) as a function of $2z/L$ 
for $N=6$ electrons confined by an external potential with $\beta=2$ and $L/a_{\rm B}=1$, 2 and 3 
in a thin wire of radius $b=0.1 a_{\rm B}$. Bottom panel: Same as in the top panel but for $\beta=6$ and $L/a_{\rm B}=2$, 4 and 6. Results for the noninteracting system are also shown in both panels for comparison.\label{fig:two}}
\end{center}
\end{figure}

\begin{figure}
\begin{center}
\includegraphics[width=1.00\linewidth]{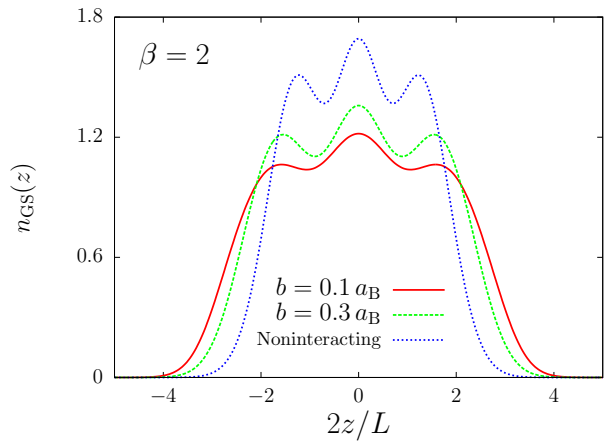}
\caption{(Color online) Density profile $n_{\rm GS}(z)$ (in units of $2/L$) as a function of $2z/L$ 
for $N=6$ electrons confined by an external potential with $\beta=2$ and $L=2 a_{\rm B}$, for two values of the wire radius. 
Results for the noninteracting system have also been included for comparison.\label{fig:three}}
\end{center}
\end{figure}

\begin{figure}
\begin{center}
\includegraphics[width=0.80\linewidth]{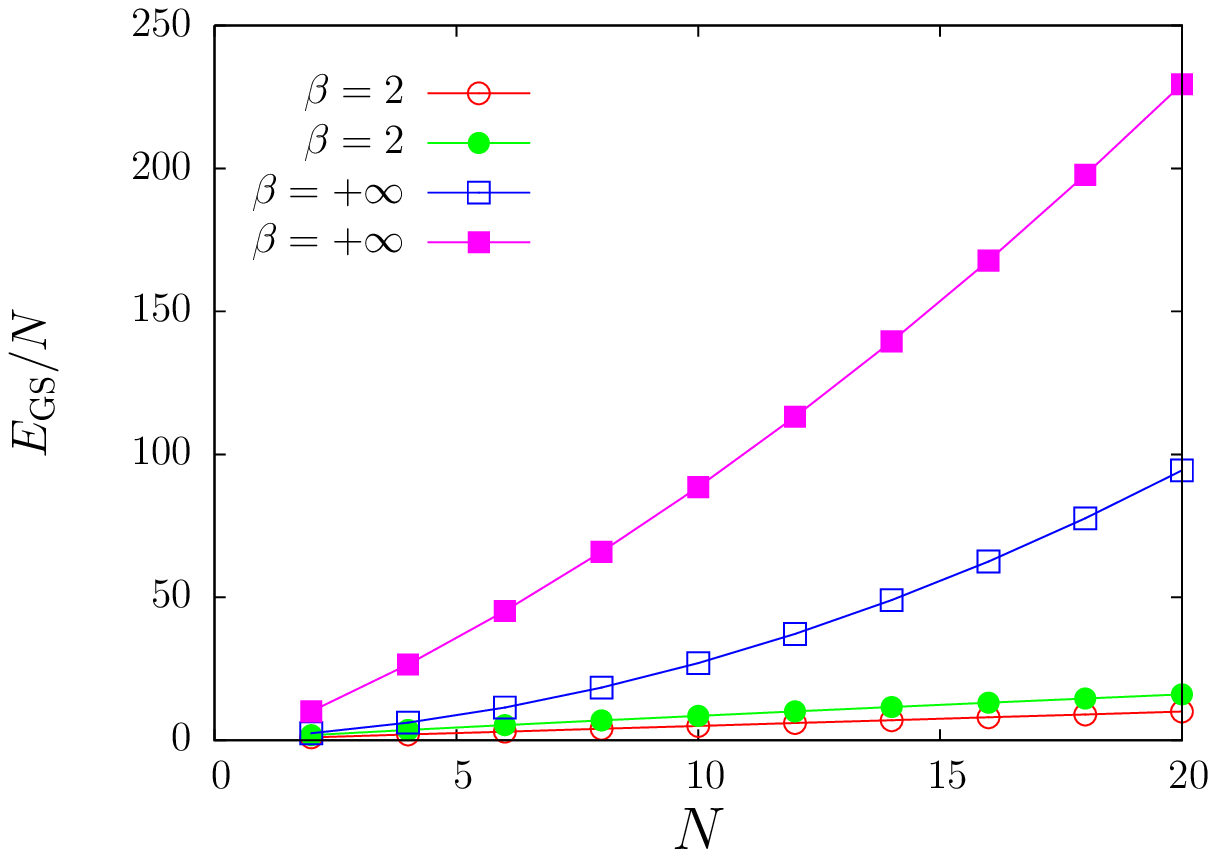}\\
\includegraphics[width=0.80\linewidth]{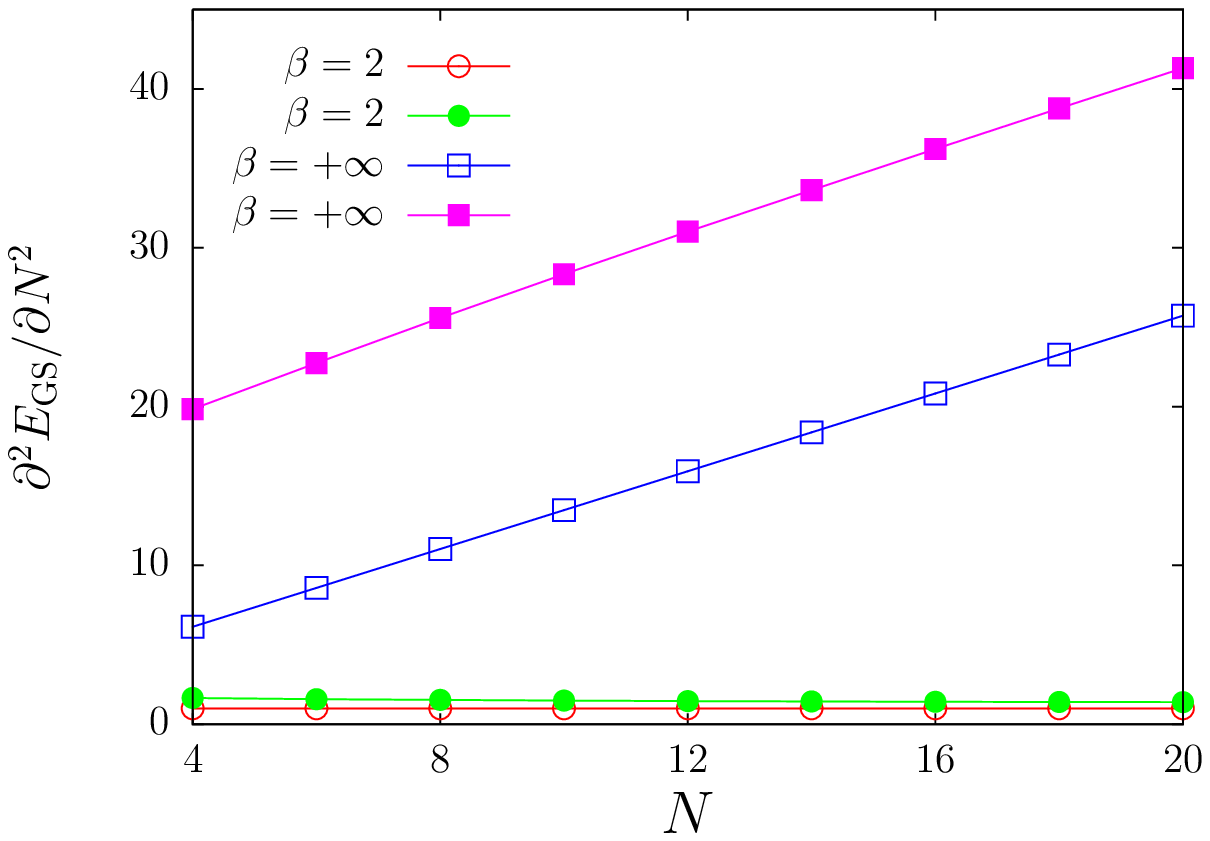}
\caption{(Color online) Top panel: Ground-state energy [per particle and in units of $2\hbar^2/(m L^2)$] as a function of the electron number $N\geq 2$ for various values of $\beta$ and $b=0.1 a_{\rm B}$. Filled symbols correspond to the interacting case, empty symbols to noninteracting case. For $\beta=2$ we have chosen $L=a_{\rm B}$, while for $\beta=+\infty$ $L=5 a_{\rm B}$. Bottom panel: Thermodynamic stiffness $\partial^2 E_{\rm GS}/\partial N^2$ [in units of $2\hbar^2/(m L^2)$] as a function of $N$ for the same system parameters as in the top panel. The lines are just guides for the eye.\label{fig:four}}
\end{center}
\end{figure}

We have solved numerically the self-consistent scheme given by Eqs.~(\ref{eq:kss})-(\ref{eq:hartree_potential}) 
using the LDA xc potential in Eq.~(\ref{eq:xc_potential}). Our main numerical results for the density profile $n_{\rm GS}(z)$ of even numbers of electrons in a weak-coupling regime are summarized in Figs.~\ref{fig:one}-\ref{fig:four}.

In the homogeneous limit the $1D$ hypothesis ({\it i.e.} a single transverse subband occupied) requires that the Fermi energy $\varepsilon_{\rm F}=\hbar^2 k^2_{\rm F}/(2m)$, with $k_{\rm F}=\pi n/2=\pi/(4 r_s a_{\rm B})$, be smaller than the transverse energy $\hbar \omega_\perp$. This translates into the inequality 
$r_s > \pi {\bar b}/4$, involving $r_s$ and the wire radius $b$ in units of the Bohr radius. In our calculations we have checked that the {\it minimum} $r_s(z)$ defined by the local density $n_{\rm GS}(z)$
satisfies the $1D$ hypothesis for each set $(N, {\bar b}, \beta, {\bar L})$ of parameters.

In Fig.~\ref{fig:one} we report the density profiles for $N=4, 6$, and $8$ electrons in the case of a thin wire with radius $b=0.1\,a_{\rm B}$ and a confinement with $L=a_{\rm B}$, corresponding to $\lambda \approx 4$. We see from this figure that for these system parameters the ground state is fluid-like with $N/2$ distinct maxima, corresponding to Friedel-like oscillations with wave number $2 k^{\rm eff}_{\rm F}$ where the effective Fermi wavenumber $k^{\rm eff}_{\rm F}=\pi {\widetilde n}/2$ is determined by the average density ${\widetilde n}$ in the bulk of the trap.
In Fig.~\ref{fig:two} we show the evolution of the density profile with increasing $L$ for $N=6$ electrons confined in a thin wire of radius 
$b=0.1\,a_{\rm B}$, and in Fig.~\ref{fig:three} we show the evolution of the density profile with increasing $b$ for fixed $L=2 a_{\rm B}$. 
The role of electron-electron interactions becomes more important with increasing $L$ or decreasing $b$ (for $L=6 a_{\rm B}$ and $b=0.1 a_{\rm B}$ for example, we have $\lambda\approx 160$) and leads to a decrease in the amplitude of the Friedel-like oscillations and to a broadening of the density profile. 

In Fig.~\ref{fig:four} we report the dependence of the ground-state energy $E_{\rm GS}$ and of the stiffness $\partial^2 E_{\rm GS}/\partial N^2=[E_{\rm GS}(N+2)+E_{\rm GS}(N-2)-2E_{\rm GS}(N)]/4$ on the electron number $N$, for different types of confining potential. The behavior of these quantities is easily understood in the noninteracting case. In harmonic confinement the single-particle spectrum is given by  $\varepsilon_i=\hbar \omega_{\|}(i+1/2)$ with $i=0,1,2,...$ and thus the ground-state energy is 
$E_{\rm GS}(N)=2\sum_{i=0}^{N/2-1}\varepsilon_i=\hbar\omega_{\|} N^2/4$, implying a constant stiffness $\partial^2 E_{\rm GS}/\partial N^2=\hbar\omega_{\|}/2$. 
In the case $\beta=+\infty$, instead, $\varepsilon_i=\hbar^2 \pi^2 i^2/(2 m L^2)$ with $i=1,2,3,...$ and thus $E_{\rm GS}(N)=\hbar^2\pi^2 N(N+1)(N+2)/(24 m L^2)$, implying a linear stiffness $\partial^2 E_{\rm GS}/\partial N^2=\hbar^2\pi^2(N+1)/(4 m L^2)$.
We are instead unable to calculate the addition energy~\cite{kleimann_2000} (chemical potential) $\mu=E_{\rm GS}(N)-E_{\rm GS}(N-1)$, as it requires knowledge of the ground-state energy for systems having odd numbers of electrons and hence a finite spin polarization. The spin-polarization dependence of the correlation energy of the $1D$ EL is presently not yet available.

Whereas the above results refer to a fluid-like weak-coupling regime, one should expect real-space quasi-ordering to set in at strong coupling, and this should be signaled by the so-called ``$2k_F\rightarrow 4k_F$ crossover'' in the wave number of Friedel oscillations. This cross-over is not predicted by the LDA xc functional in Eq.~(\ref{eq:xc_potential}). In Section~\ref{sect:strong_coupling} we study in detail this crossover for $N=2$ harmonically-trapped electrons, a problem which is easily solvable numerically to any desired degree of accuracy (see also the work of Szafran {\it et al.}~\cite{bednarek_2003}).

\section{The two-particle problem and the failure of the LDA at strong coupling}
\label{sect:strong_coupling}

\begin{figure}
\begin{center}
\includegraphics[width=1.00\linewidth]{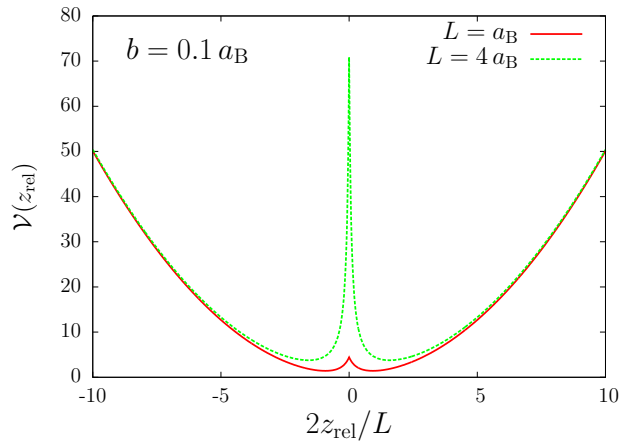}
\caption{(Color online) The effective potential ${\cal V}(z_{\rm rel})$ [in units of $2\hbar^2/(m L^2)$] as a function of $2z_{\rm rel}/L$ for $b=0.1 a_{\rm B}$.\label{fig:five}}
\end{center}
\end{figure}

After a canonical transformation to centre-of-mass and relative coordinates and momenta [$Z=(z_1+z_2)/2,P=p_1+p_2$ and $z_{\rm rel}=z_1-z_2,p=(p_1-p_2)/2$], the Hamiltonian for two harmonically trapped electrons in a thin wire can be written as
${\cal H}={\cal H}_{\rm CM}(Z,P)+{\cal H}_{\rm rel}(z_{\rm rel},p)$. Here, 
the centre-of-mass Hamiltonian ${\cal H}_{\rm CM}=P^2/(2M)+M\omega^2_{\|}Z^2/2$ describes 
a $1D$ harmonic oscillator of mass $M=2m$, while the relative-motion Hamiltonian ${\cal H}_{\rm rel}=p^2/m+{\cal V}(z_{\rm rel})$ 
describes a particle of mass $m/2$ in the potential ${\cal V}(z_{\rm rel})=m\omega^2_{\|}z^2_{\rm rel}/4+v_b(z_{\rm rel})$. This potential is plotted in Fig.~\ref{fig:five} for two values of the trap frequency $\omega_{\|}=4\hbar/(m L^2)$. 

\begin{figure*}
\begin{center}
\tabcolsep=0cm
\begin{tabular}{cc}
\includegraphics[width=0.45\linewidth]{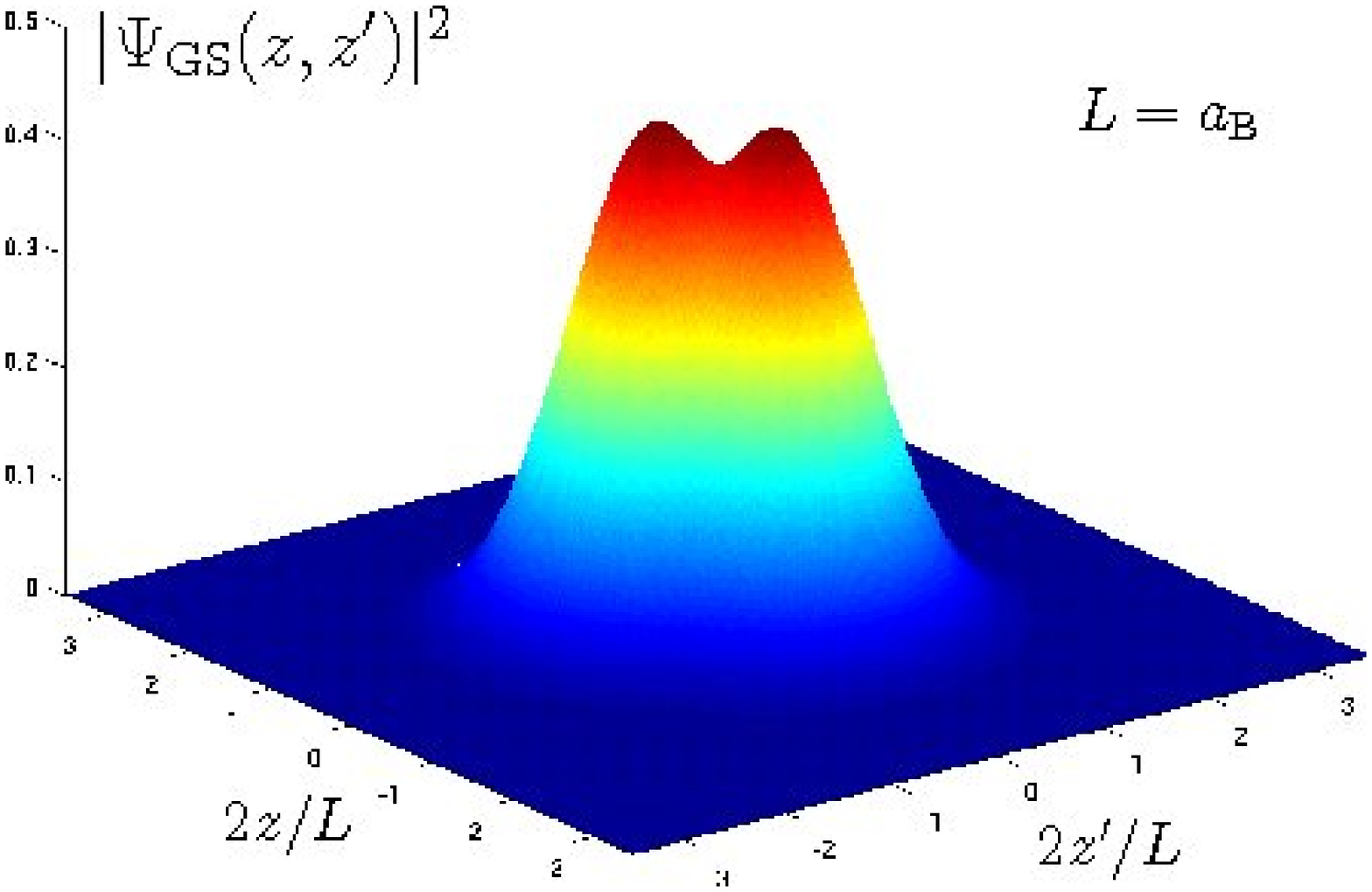} &\hspace{0.6 cm} \includegraphics[width=0.45\linewidth]{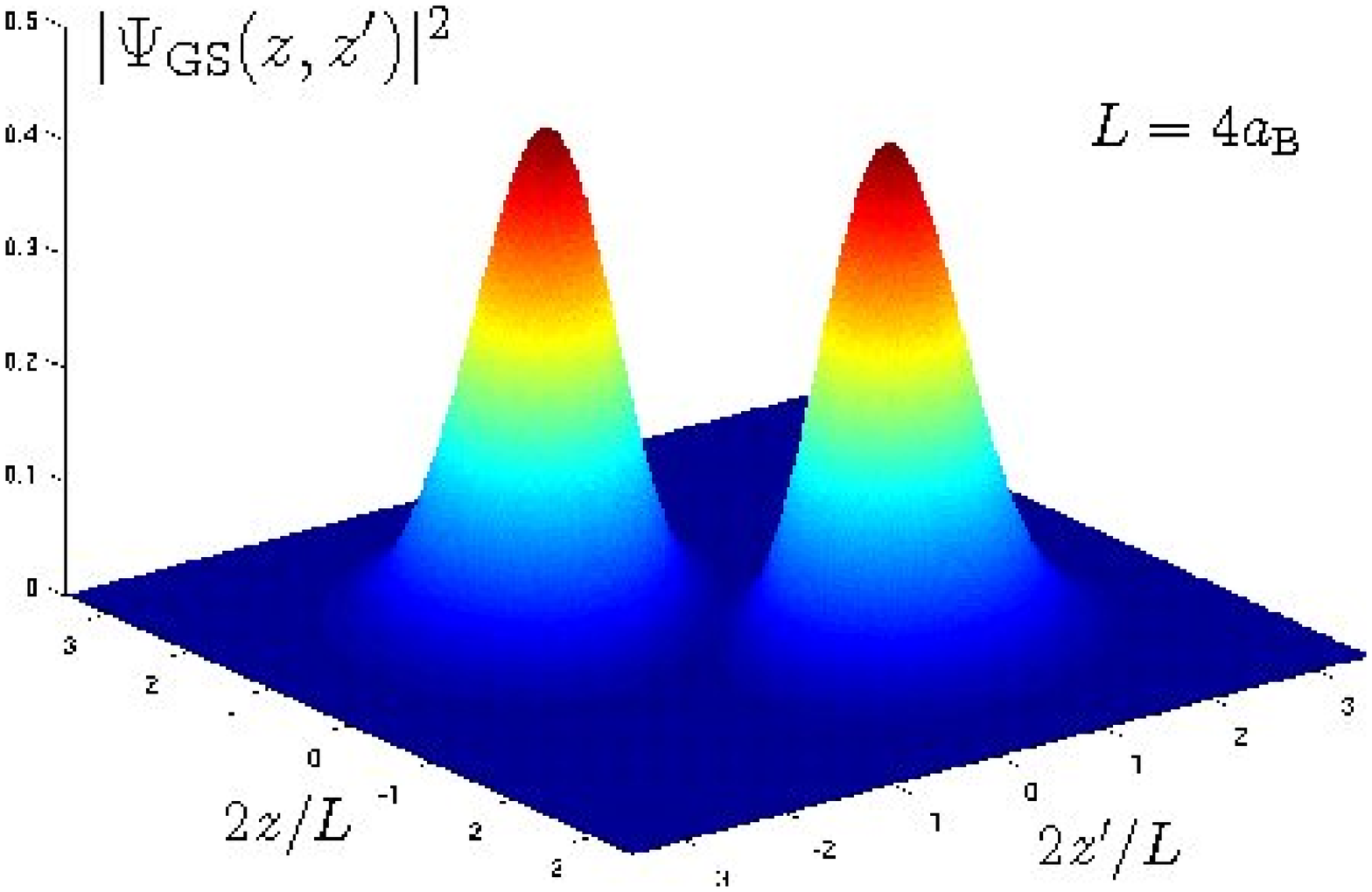}
\end{tabular}
\caption{(Color online) Exact two-body wavefunction $|\Psi_{\rm GS}(z,z')|^2$ [in units of $(2/L)^2$] as a function of $2z/L$ and $2z'/L$
for $N=2$ electrons confined by a harmonic potential with $L=a_{\rm B}$ (left panel) and $L= 4 a_{\rm B}$ (right panel) 
in a thin wire of radius $b=0.1 a_{\rm B}$.\label{fig:six}}
\end{center}
\end{figure*}

In the spin-singlet case the spatial part of the ground-state wavefunction is written as
\begin{equation}\label{eq:two_body}
\Psi_{\rm GS}(z_1,z_2)={\cal N}\exp{(-Z^2/a_{\|}^2)}\varphi_{\rm rel}(z_{\rm rel})\,,
\end{equation}
where ${\cal N}$ is a normalization constant, $a_{\|}=\sqrt{\hbar/(m\omega_{\|})}$, and $\varphi_{\rm rel}(z_{\rm rel})$ is the symmetric ground-state wavefunction for the relative-motion problem with energy $\varepsilon_{\rm r}$, which can be numerically found by solving the single-particle Schr\"odinger equation
\begin{equation}
\left[-\frac{\hbar^2}{m}\frac{d^2}{d z_{\rm rel}^2}+{\cal V}(z_{\rm rel})\right]\varphi_{\rm rel}(z_{\rm rel})=\varepsilon_{\rm r}
\varphi_{\rm rel}(z_{\rm rel})\,.
\end{equation}
An illustration of $|\Psi_{\rm GS}(z,z')|^2$ for two values of ${\bar L}$ is reported in Fig.~\ref{fig:six}. The ``molecular" nature of the ground state is evident at strong coupling.

\begin{figure*}
\begin{center}
\tabcolsep=0 cm
\begin{tabular}{cc}
\includegraphics[width=0.50\linewidth]{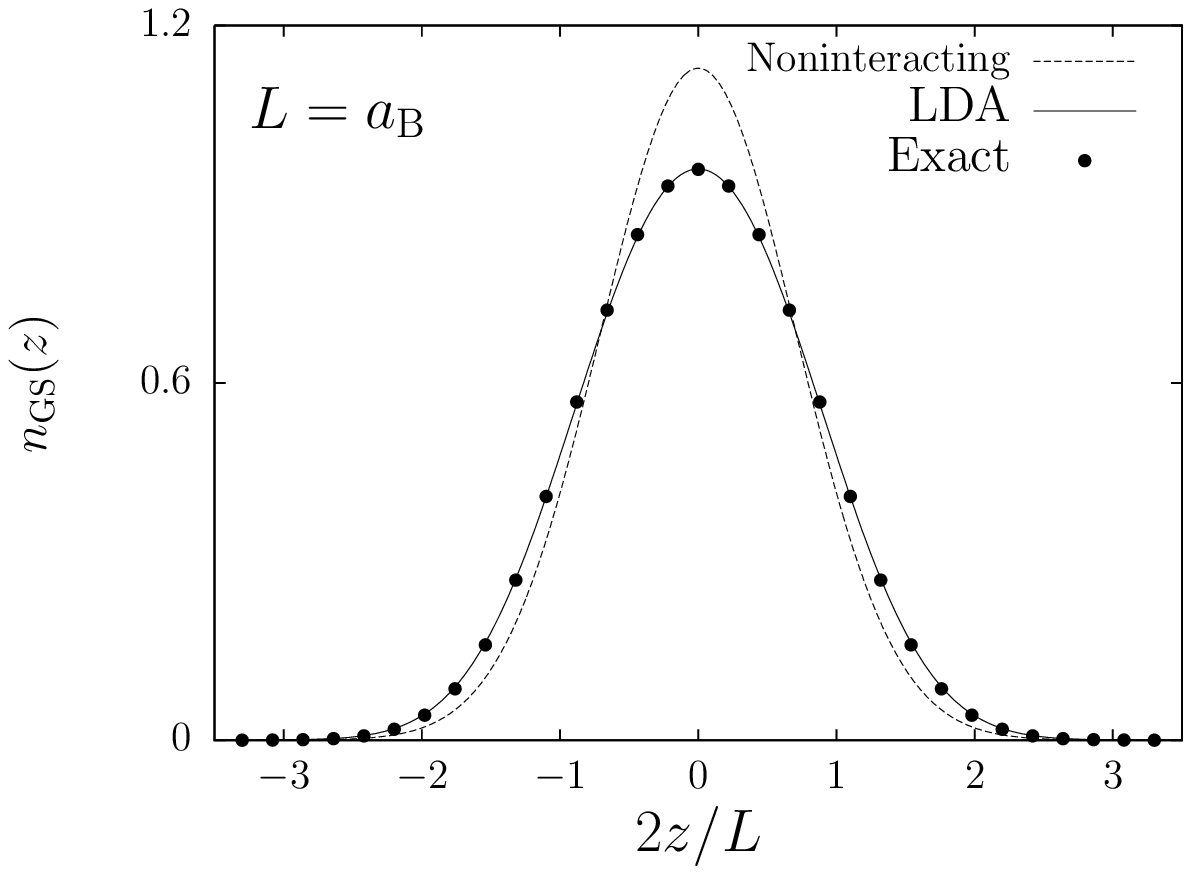}&
\includegraphics[width=0.50\linewidth]{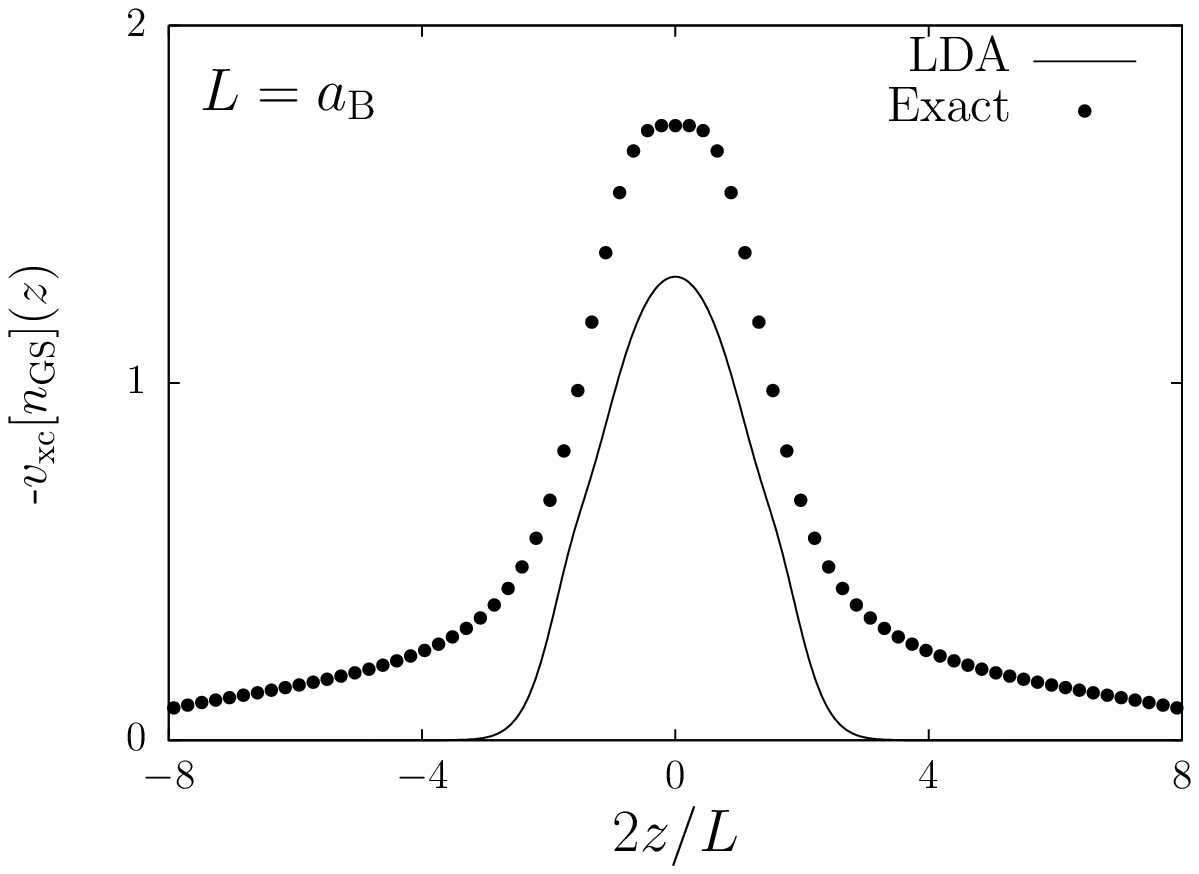}\\
\includegraphics[width=0.50\linewidth]{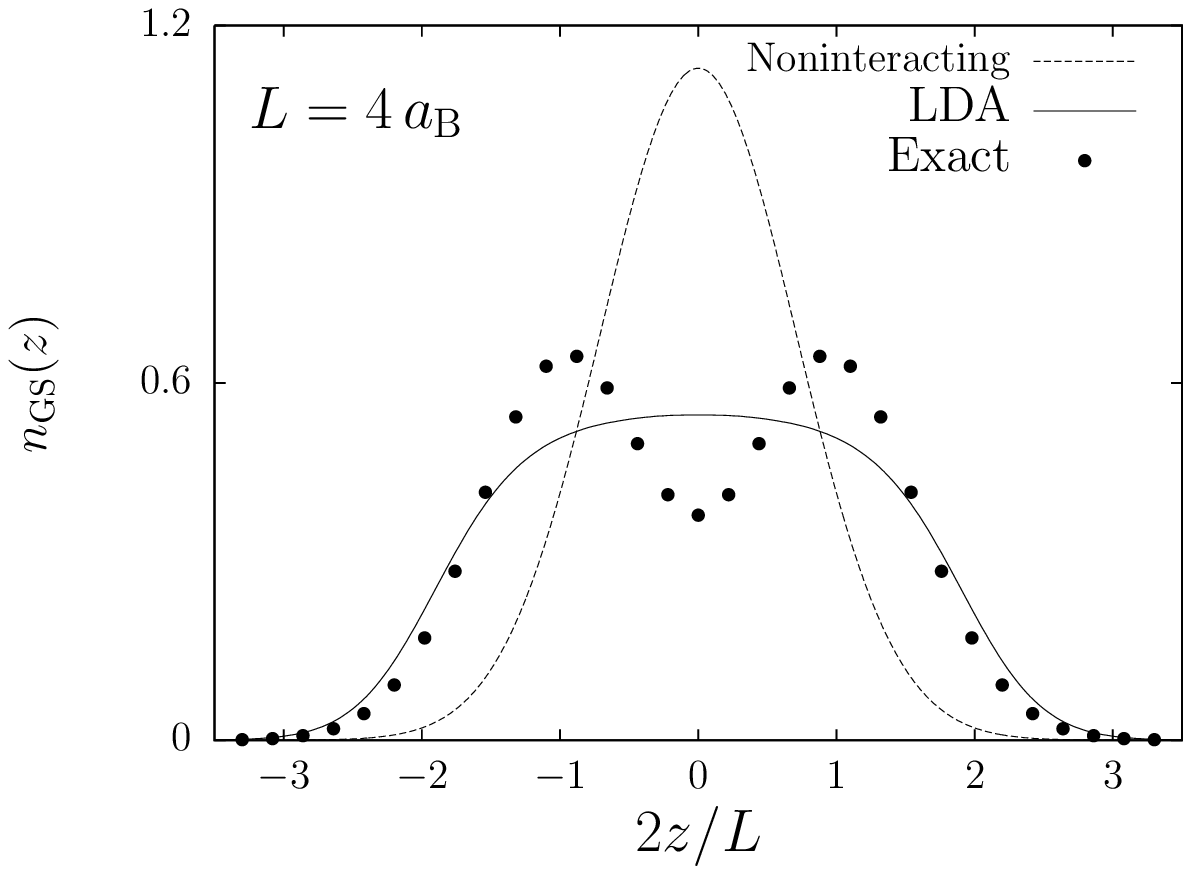}&
\includegraphics[width=0.50\linewidth]{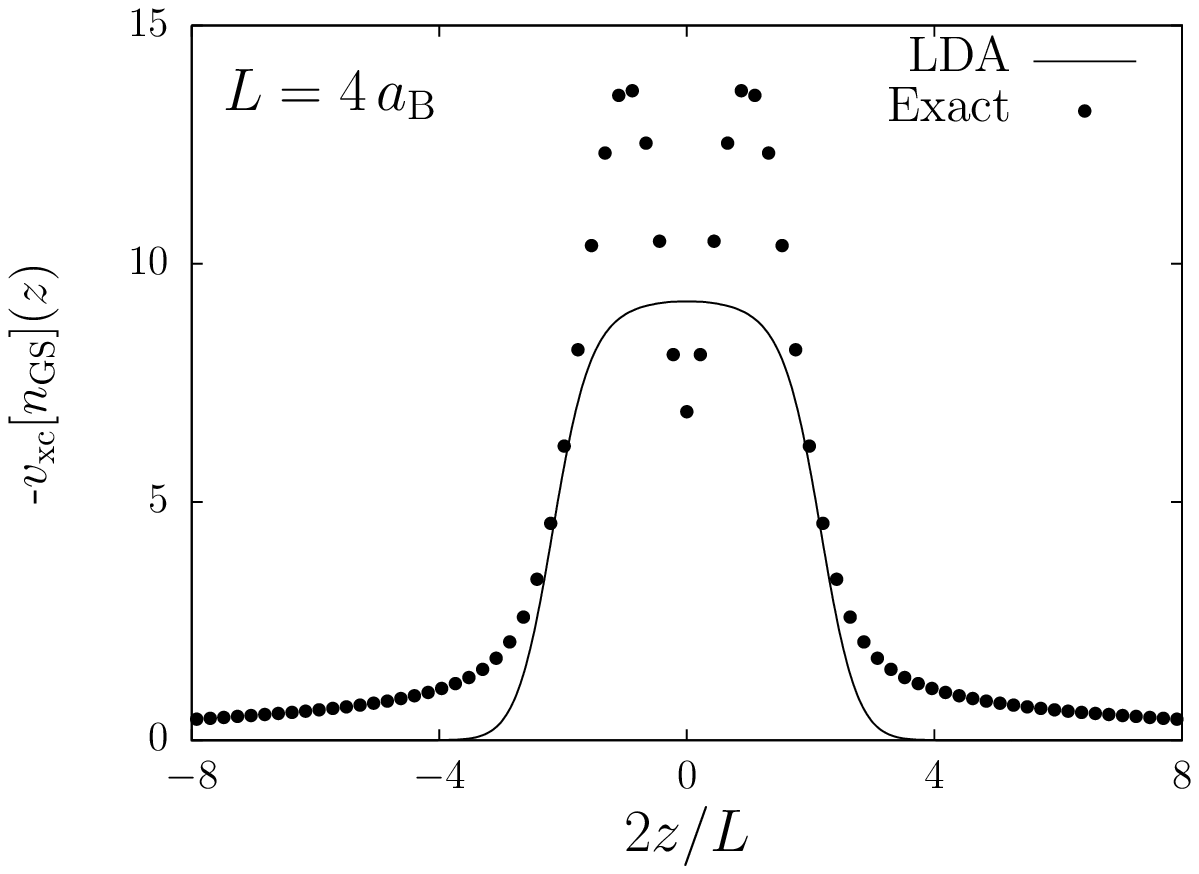}
\end{tabular}
\caption{Left panels: Density profile $n_{\rm GS}(z)$ (in units of $2/L$) as a function of $2z/L$ 
for $N=2$ electrons confined by a harmonic potential with $L=a_{\rm B}$ (top) and $L= 4 a_{\rm B}$ (bottom) 
in a thin wire of radius $b=0.1 a_{\rm B}$. The exact results (filled circles) are compared with the LDA results (solid line).
Right panels: The LDA xc potential from Eq.~(\ref{eq:xc_potential}) (solid line)
is compared with the exact xc potential calculated from Eq.~(\ref{eq:exact_vxc}) (filled circles).\label{fig:seven}}
\end{center}
\end{figure*}

The ground-state density profile can be found from 
\begin{equation}\label{eq:exact_density}
n_{\rm GS}(z)=\int_{-\infty}^{+\infty}dz'\,|\Psi_{\rm GS}(z,z')|^2\,,
\end{equation}
where the normalization constant ${\cal N}$ is chosen according to $\int_{-\infty}^{+\infty}dz\,n_{\rm GS}(z)=2$.
Numerical results are shown in Fig.~\ref{fig:seven} (left panels) in comparison with the LDA profiles. Note that the double-peak structure in $|\Psi_{\rm GS}(z,z')|^2$ at weak coupling (left panel in Fig.~\ref{fig:six}) is lost in the corresponding ground-state density. While at weak coupling ($L= a_{\rm B}$) the agreement between the exact  result and the LDA prediction is very satisfactory, at strong coupling ($L=4 a_{\rm B}$) the LDA is unable to reproduce the formation of a deep Coulomb hole yielding a density profile with a broad maximum at the trap center. 

One can also directly compare the LDA xc potential in Eq.~(\ref{eq:xc_potential}) with the exact one, which can 
be calculated from the exact density profile~\cite{laufer} as summarized below.
In the two-particle case there is only {\it one} Kohn-Sham orbital 
$\varphi_{\rm KS}(z)=\sqrt{n_{\rm GS}(z)/2}$, which satisfies the Kohn-Sham equation
\begin{equation}
\left[-\frac{\hbar^2}{2m}\frac{d^2}{d z^2}+V_{\rm KS}[n_{\rm GS}](z)\right]\varphi_{\rm KS}(z)=\varepsilon_{\rm KS}\varphi_{\rm KS}(z)\,.
\end{equation}
Solving this equation for $v_{\rm xc}$ we find
\begin{eqnarray}
v_{\rm xc}[n_{\rm GS}](z)&=&\varepsilon_{\rm KS}+\frac{\hbar^2}{2m\varphi_{\rm KS}(z)}\frac{d^2\varphi_{\rm KS}(z)}{d z^2}-V_{\rm ext}(z)\nonumber\\
&-&v_{\rm H}(z)\,,
\end{eqnarray}
or, more explicitly,
\begin{eqnarray}\label{eq:exact_vxc}
v_{\rm xc}[n_{\rm GS}](z)&=&\varepsilon_{\rm KS}+\frac{\hbar^2}{2m\sqrt{n_{\rm GS}(z)}}\frac{d^2 \sqrt{n_{\rm GS}(z)}}{d z^2}-V_{\rm ext}(z)\nonumber\\
&-&\int_{-\infty}^{+\infty}dz'~v_b(|z-z'|)n_{\rm GS}(z')\,.
\end{eqnarray}
The exact Kohn-Sham eigenvalue $\varepsilon_{\rm KS}$ can be proven to be equal to the energy $\varepsilon_{\rm r}$ of the relative motion. The approximate Kohn-Sham eigenvalue, instead, differs from $\varepsilon_{\rm r}$: for example, for $L=a_{\rm B}$ we find $\delta\equiv \varepsilon^{\rm LDA}_{\rm KS}-\varepsilon_{\rm r}\simeq 0.46~[2\hbar^2/(m L^2)]$.

In Fig.~\ref{fig:seven} (right panels) we show a comparison between the LDA xc potential in Eq.~(\ref{eq:xc_potential}), as obtained at the end of the Kohn-Sham self-consistent procedure, and the exact xc potential calculated from Eq.~(\ref{eq:exact_vxc}) with the ground-state density from Eq.~(\ref{eq:exact_density}). Several remarks are in order here. As it commonly happens, the LDA potential has the wrong long-distance behavior: it decays exponentially because the density does so, while the exact xc potential decays like $1/|z|$. Nevertheless, at weak coupling the difference between the two potentials is well approximated by the constant $\delta$, in the region where the density profile is different from zero, and this explains the satisfactory agreement between the exact and the LDA profiles. It is finally evident how in the strong-coupling regime the LDA potential is instead very different from the exact one.

An xc functional that embodies the $2k_{\rm F}\rightarrow 4k_{\rm F}$ crossover and is capable of describing inhomogeneous 
Luttinger systems at strong repulsive coupling is thus required. In Ref.~\onlinecite{saeed_pra_2006} we have proposed a simple xc functional which is able to capture the tendency to antiferromagnetic spin ordering. 
The idea consists in two steps: (i) one adds an infinitesimal spin-symmetry-breaking field to the Hamiltonian; and (ii) one resorts to a local spin-density approximation (LSDA) within the framework of spin-density functional theory. Earlier exact diagonalization and configuration-interaction studies of $1D$ quantum dots~\cite{jauregui_1993,bednarek_2003} have shown that, while for even number of electrons the local spin polarization is everywhere zero in the dot, one can still observe antiferromagnetic correlations at strong coupling by looking at the spin-resolved pair correlation functions. 
This suggests that an LSDA approach may indeed prove useful at strong coupling. 
Unfortunately, a knowledge of the ground-state energy of the homogeneous $1D$ EL in the situations with $N_\uparrow \neq N_\downarrow$ is still lacking.

\section{Conclusions}
\label{sect:discussion_conclusions}

In summary, we have carried out a novel density-functional study of a few isolated electrons at zero net spin, 
confined by power-law external potentials inside a short portion of a thin semiconductor quantum wire. 
The theory employs the quasi-one-dimensional homogeneous electron liquid as the reference system 
and transfers its ground-state correlations to the confined inhomogeneous system 
through a local-density approximation to the exchange and correlation energy functional. 

The local-density approximation gives good-quality results for the density profile in the liquid-like states of the system at weak coupling, a precise test against exact results having been presented in the case of $N=2$ electrons. However, it fails to describe the emergence of electron localization into Wigner molecules at strong coupling. The fact that strong-coupling antiferromagnetic correlations are hidden in the inner-coordinates degrees of freedom, as suggested by Szafran {\it et al.}~\cite{bednarek_2003}, indicates that  a local spin-density approximation, or even non-local functionals based on the spin-resolved pair correlation functions~\cite{gunnarsson_1979}, are needed. The class of density-functional schemes for ``strictly correlated'' electronic systems recently proposed by Perdew {\it et al.}~\cite{seidl_1999} may also be useful in treating the Wigner-molecule regime.

\acknowledgments

We are indebted to M. Casula for providing us with 
his QMC data prior to publication. It is a pleasure to thank R. Asgari, K. Capelle, P. Capuzzi, M. Governale, 
I. Tokatly, and G. Vignale for several useful discussions.

\end{document}